\documentclass[twocolumn,showpacs,preprintnumbers,prb,amsmath,amssymb]{revtex4-1}

\usepackage{graphicx}
\usepackage{dcolumn}
\usepackage{bm}

\begin{document}


\title{Compact wavelength add-drop multiplexers using Bragg gratings in coupled dielectric-loaded plasmonic waveguides}

\author{G. Biagi,$^{1,2}$ J.~Fiutowski,$^3$ I.~P.~Radko,$^{2,4,*}$ H.-G.~Rubahn,$^3$ K.~Pedersen,$^1$ and S.~I.~Bozhevolnyi$^2$}

\affiliation{
$^1$Department of Physics and Nanotechnology, Aalborg University, Skjernvej 4A, 9220 Aalborg \O{}st, Denmark\\
$^2$Centre for Nano Optics, University of Southern Denmark, Niels Bohrs All\'e 1, 5230 Odense M, Denmark\\
$^3$Mads Clausen Institute, NanoSYD, University of Southern Denmark, Alsion 2, 6400 S\o{}nderborg, Denmark\\
$^4$Department of Physics, Technical University of Denmark, 2800 Kongens Lyngby, Denmark
}
\email{Corresponding author: iradko@fysik.dtu.dk}

\date{\today}

\begin{abstract}
We report a novel design of a compact wavelength add-drop multiplexer utilizing dielectric-loaded surface plasmon-polariton waveguides (DLSPPWs). The DLSPPW-based configuration exploits routing properties of directional couplers and filtering abilities of Bragg gratings. We present practical realization of a 20-$\mu$m-long device operating at telecom wavelengths that can reroute optical signals separated by approximately 70~nm in the wavelength band. We characterize the performance of the fabricated structures using scanning near-field optical microscopy as well as leakage-radiation microscopy and support our findings with numerical simulations.
\end{abstract}


\maketitle

\noindent Surface-plasmon circuitry~\cite{1} numbers a large variety of waveguiding configurations, ranging from tightly confined and very lossy surface plasmon-polariton (SPP) modes in metal nanowires~\cite{2} and chains of nanospheres~\cite{3} to moderate and long-travelling plasmonic modes, such as channel plasmon polaritons~\cite{4}, dielectric-loaded SPP waveguide (DLSPPW) modes~\cite{5}, long-range SPPs~\cite{6}, and long-range DLSPPW modes~\cite{7}.

Considering only propagation loss and confinement characteristics, there is a technology superior to plasmonics, viz., silicon photonics. However, metal stripes as an inherent part of plasmonic waveguides provide easy access to the external electrodes. Moreover, the fact that the SPP fields reach their maximum at the metal--dielectric interface makes electro- and thermo-optic modulation substantially more energetically efficient than in the case of silicon photonics~\cite{13,14,15}, since controlling electrodes can be placed right in the center of the SPP mode.

Among all plasmonic configurations, DLSPPWs are so far the only waveguides that have been exploited for realization of complex circuit elements~\cite{8,9}, active control~\cite{10}, and partial loss compensation using optical pumping~\cite{11,12}. Furthermore, a low-energy thermo-optical tuning and fast error-free 10-Gb/s transmission through DLSPPW-based components has been recently demonstrated~\cite{15}, paving the way towards practical telecom applications. Importantly, DLSPPWs are naturally compatible with industrial fabrication using large-scale ultra-violet~(UV) lithography~\cite{8}.

While a great deal of DLSPPW-based passive and active photonic components has already been demonstrated, the task of spatial separation of telecom channels at different wavelengths appeared to be rather challenging due to considerable propagation losses in DLSPPWs~\cite{5}. Thus, configurations based on waveguide ring or racetrack resonators that have to employ sufficiently long resonators to minimize bend losses feature typically low transmission levels~\cite{15}. In this work, we present a very compact design of a wavelength add-drop multiplexer (WADM) based on DLSPPWs, which exploits filtering abilities of Bragg gratings~\cite{8} and routing properties of directional couplers~\cite{9}. The proposed configuration involves thereby only \textit{straight} (coupled) waveguides that should be long enough to facilitate efficient Bragg reflection and power transfer --- the requirements that can, in principle, be met with short ($<20~\mu$m) DLSPPWs~\cite{8,9}, implying absorption losses below 2~dB when operating at telecom wavelengths~\cite{5}.

\begin{figure}[tbhp]
\centerline{\includegraphics[width=0.8\columnwidth]{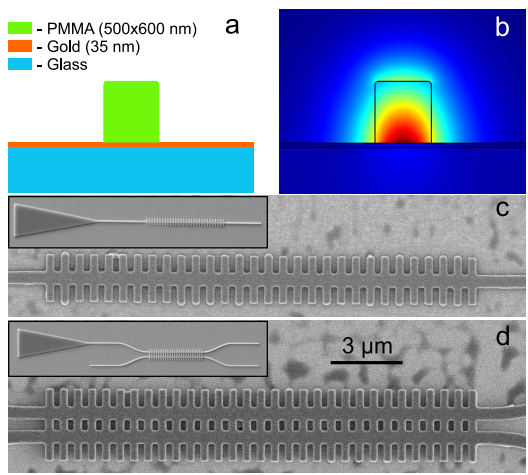}}
\caption{\label{fig1}
a)~Schematic cross section of the sample with a single DLSPPW on top. b)~Finite-element modelling of electric-field distribution inside the DLSPPW. c)~and d)~SEM micrographs of the active part of the fabricated c)~DLSPPW-based Bragg grating and d)~wavelength add-drop multiplexer (WADM). Insets show the design of the whole structure. The DLSPPW ridges have dimensions of 500~nm in width and 600~nm in height. The transverse ridges of the Bragg gratings in both structures are 300~nm in width and repeated with a period of $\Lambda=600$~nm. For the DLSPPW Bragg grating, the length of reflecting ridges is 2.0~$\mu$m, whereas for the WADM, they are 3.2~$\mu$m in length. The length of the coupling region in WADM is 20.0~$\mu$m. Center-to-center separation between DLSPPWs in the coupling region of the WADM is 1.2~$\mu$m.}
\end{figure}

The main and original idea behind our WADM design relies on realization of \textit{concomitant} power transfer (by directional coupling) from the input to the parallel (add-drop) waveguide and Bragg filtering (reflection) of the transmitted (rejected) mode by a Bragg grating superimposed with both waveguides. For this reason, our work comprises fabrication and characterization of Bragg gratings as well as WADMs based on DLSPPWs (Fig.~\ref{fig1}).

We used e-beam lithography to fabricate DLSPPW structures on a transparent glass substrate covered with a 35-nm-thick gold film with a 2-nm Ti adhesion layer underneath gold [Fig.~\ref{fig1}(a)]. We exploited the previously found for telecom wavelengths optimal DLSPPW geometry of 500~nm in width and 600~nm in height~\cite{5}, which provides subwavelength mode confinement [Fig.~\ref{fig1}(b)] without introducing much of additional loss. We further checked actual dimensions of the fabricated DLSPPWs with atomic-force (AFM) and scanning electron (SEM) microscopes [Figs.~\ref{fig1}(c),(d)].

To test the operation properties of the fabricated waveguides, we exploited scanning near-field optical microscope (SNOM) operating in collection mode and leakage-radiation microscopy (LRM)~\cite{16}. In the former case, we used an uncoated tapered optical fiber as a SNOM probe, whereas for excitation of DLSPPW modes we employed standard Kretschmann configuration with the excitation angle adjusted to the DLSPPW mode effective index~\cite{17}. In case of LRM, for DLSPPW mode excitation, we illuminated DLSPPW tapers [cf. insets in Figs.~\ref{fig1}(c),(d)] with a normally incident focused laser beam~\cite{16}. In both SNOM and LRM setups a tunable laser emitting in the telecommunication wavelength range ($\lambda=1500~\text{nm}-1640~\text{nm}$) served as a source of light.

\begin{figure}[tbhp]
\centerline{\includegraphics[width=\columnwidth]{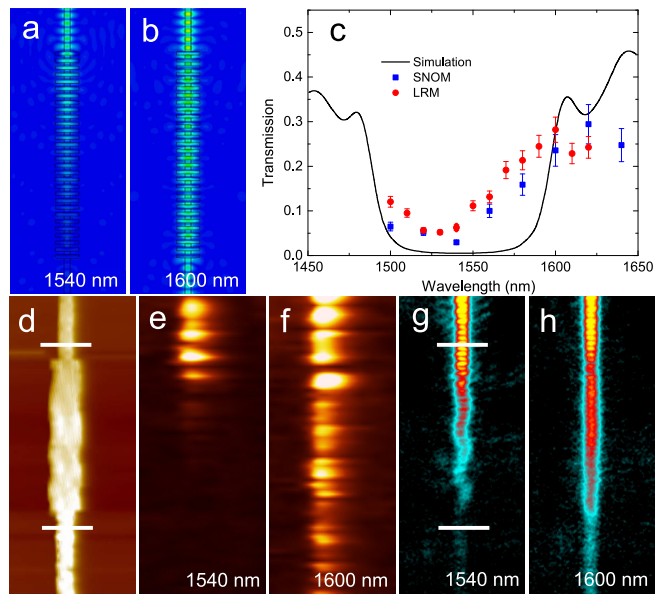}}
\caption{\label{fig2}
a)~and b)~Simulated electric-field plots of DLSPPW mode propagating through the Bragg grating at a)~$\lambda=1540$~nm and b)~$\lambda=1600$~nm. c)~Intensity transmission curve of the DLSPPW Bragg grating: simulations and experimental results. d)~SNOM topography of the DLSPPW Bragg grating. e), f)~SNOM and g), h)~LRM optical images (intensity) of DLSPPW mode propagating through the Bragg grating acquired at e), g)~$\lambda=1540$~nm and f), h)~$\lambda=1600$~nm. Experimental SNOM and LRM images are to the same scale. White bars show positions where in and out signals where taken for evaluation of the transmission.}
\end{figure}

First, we investigated several straight DLSPPWs with Bragg gratings. To facilitate the choice of geometrical parameters, prior to conducting experimental characterization of any structure, we carried out finite-difference time-domain simulations (Lumerical, Lumerical Solutions, Inc.). Experimental investigations confirmed the results of numerical modelling predicting well-pronounced wavelength filtering of the DLSPPW-based Bragg grating (Fig.~\ref{fig2}). For each case --- simulations [Figs.~\ref{fig2}(a),(b)], SNOM [Figs.~\ref{fig2}(d)-(f)], and LRM [Figs.~\ref{fig2}(g),(h)] measurements --- a pair of images with maximum and minimum transmission clearly demonstrates the filtering properties of the structure. For the grating period of $\Lambda=600$~nm, the mode is almost completely reflected at wavelengths around 1540~nm. With an increase of the wavelength, the transmission of the DLSPPW mode through the grating increases significantly.

The number of grating periods was fixed to $N=30$ [Fig.~\ref{fig1}(c)], which appeared to be a good trade-off between the filtering ability and insertion loss caused primarily by the ohmic loss (radiation absorption)~\cite{8}. For our system parameters, the length of Bragg grating amounts to 18~$\mu$m, so that the absorption loss should be below 2~dB~\cite{5}, leaving slightly more than 1~dB to additional (scattering) loss. We expect that the latter loss can further be decreased by making use of dedicated design of the grating profile. Here, it should be noted that the Bragg grating filter is a well-known component in plasmon-based integrated optics with different approaches to structural optimization being developed~\cite{18,19}.

By taking averaged cross sections through experimental SNOM [Figs.~\ref{fig2}(d)-(f)] and LRM [Figs.~\ref{fig2}(g),(h)] images at the input and output of the grating, the transmission values of DLSPPW mode through the Bragg grating can be evaluated. One can see that the simulated transmission minimum at the wavelength interval between $\lambda=1500$~nm and 1590~nm agrees well with the experiment [Fig.~\ref{fig2}(c)]. Outside the transmission minimum, the DLSPPW mode transmission reaches the level of $\sim0.5$ (3~dB loss) and 0.3 (5~dB loss) in simulations and experiment, respectively. We found also that the ratio of maximum to minimum transmission through the EBL-fabricated DLSPPW Bragg grating exceeds that for UV lithography-fabricated structure (data not shown). This implies high sensitivity of the structure efficiency to the fabrication quality of its nanometer-sized features, most probably due to additional scattering by fabrication-induced irregularities.

\begin{figure}[tbhp]
\centerline{\includegraphics[width=0.8\columnwidth]{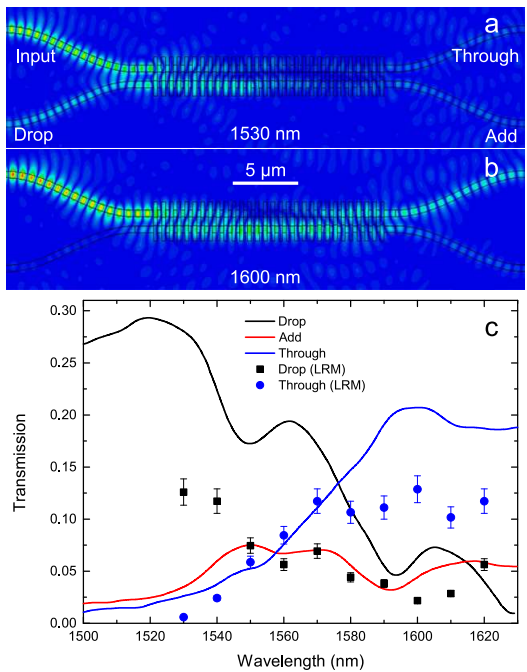}}
\caption{\label{fig3}
Simulated electric-field plots of DLSPPW mode propagating through the WADM at a)~$\lambda=1530$~nm and b)~$\lambda=1600$~nm. c)~Comparison of normalized DLSPPW mode intensity at ``Add'', ``Drop'', and ``Through'' ports obtained in simulations and LRM experiment. Geometrical mode path length between ``Input'' and ``Through'' terminals where the signal values were taken for evaluation is $\sim40~\mu$m, which corresponds to left and right image borders on panels a) and~b).}
\end{figure}

After characterizing the simple DLSPPW-based Bragg filter/reflector, we could work out the design of DLSPPW-based WADM [Fig.~\ref{fig1}(d)], which is essentially a directional coupler combined with a Bragg filter/reflector. The DLSPPW mode at the wavelength of minimum transmission will be reflected back by the Bragg grating and simultaneously coupled to the adjacent waveguide. If the coupling efficiency (per length of propagation) matches the reflection efficiency, the reflected wavelength band will be dropped through the ``Drop'' port of the WADM in the lower left corner [Fig.~\ref{fig3}(a)]. At the same time, provided that the length of the coupling region is properly adjusted, the DLSPPW mode at a wavelength outside of the reflection band will couple to the adjacent DLSPPW and back during its propagation through the coupling region and will exit through the ``Through'' port of the WADM [Fig.~\ref{fig3}(b)]. Launching the signal to the ``Add'' port at a wavelength within the minimum transmission of the Bragg grating will redirect it to the ``Through'' port of the WADM, due to the symmetry of those ports relative to the ``Input'' and ``Drop'' terminals. Thus, this device can add (drop) a narrow-band signal to (from) an input high-bandwidth data stream. Note that we can tune the coupling efficiency of the directional coupler by changing the waveguide separation, whereas reflection efficiency of the Bragg grating depends on particular ridge geometry~\cite{18,19}. We find the working parameters using numerical simulations, and check them afterwards experimentally.

One can see from the simulation results [Fig.~\ref{fig3}(a)] that at the wavelength $\lambda=1530$~nm, the DLSPPW mode couples to the lower waveguide and, in agreement with the Bragg grating results shown in Fig.~\ref{fig2}, gets reflected by the structure. The output signal detected in the ``Drop'' port is therefore at its maximum, while the signal in the ``Through'' port stays below 5\% of the input power. At the wavelength $\lambda=1600$~nm the DLSPPW mode propagates through the grating without reflection [Fig.~\ref{fig3}(b)], and most of the power is registered in the ``Through'' port. At the same time, the signal detected in the ``Drop'' port reaches its local minimum. Finally, one can also notice that the simulated signal intensity in the ``Add'' port [red curve in Fig.~\ref{fig3}(c)] remains low in the entire wavelength range.

\begin{figure}[tbhp]
\centerline{\includegraphics[width=0.8\columnwidth]{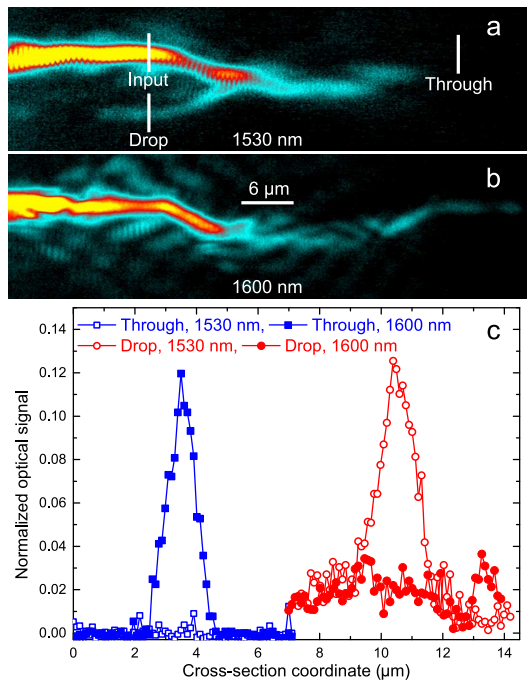}}
\caption{\label{fig4}
LRM images of DLSPPW mode propagating through the WADM at a)~$\lambda=1530$~nm and b)~$\lambda=1600$~nm. Geometrical mode path length between ``Input''/``Drop'' and ``Through'' cross-sections where the signal was taken for evaluation of transmission is $\sim40~\mu$m. c)~Normalized (to the input) intensity profiles at the ``Through'' and ``Drop'' ports at different wavelengths.}
\end{figure}

One should bear in mind that for the purpose of experimental characterization, we fabricated all DLSPPWs on top of a very thin gold film (35~nm), which allows Kretschmann mode excitation in SNOM measurements and leakage-radiation detection in LRM. Such thin film introduces additional DLSPPW mode loss due to leakage of radiation into the substrate, and for the selected wavelength range and gold-film thickness, leakage-radiation loss equals approximately to the mode ohmic loss, i.e. effectively doubles that~\cite{16}. However, if one excludes leakage loss from the system (e.g., by increasing gold thickness), the insertion loss of the WADM will roughly be equal to the ohmic loss of a straight DLSPPW section of the same length as WADM structure and is expected to be in the range of 2--4~dB.

We checked our numerical findings in the experiment using LRM (Fig.~\ref{fig4}). To characterize the spectral property of the WADM, we took averaged cross sections through the LRM images and plotted intensity distributions in the ``Drop'' and ``Through'' ports of the device [Fig.~\ref{fig4}(c)]. We normalized profiles to the intensity of the signal in the ``Input'' port. With the change of the wavelength from $\lambda=1530$~nm to 1600~nm, the WADM switches the input signal from the ``Drop'' to the ``Through'' channel. The signal in the ``Add'' port was at the noise level and therefore is not shown.

Comparison of experimental LRM and calculated normalized intensities of the DLSPPW mode in different ports of the WADM within the whole wavelength range shows qualitative agreement [Fig.~\ref{fig3}(c)] of the data. The maximum signal intensity in the experiment does not reach as high values as in simulations due to scattering on imperfections of the fabricated structure. However, both in simulations and in experiment, the two curves for the ``Drop'' and ``Through'' channels intersect at $\lambda\approx1570$~nm, switching the signal from one port to another. We stress once again, that if one excludes the purposefully introduced leakage loss from the system, the maximum values in ``Drop'' and ``Through'' ports will roughly double, thus increasing the maximum to minimum transmission ratio and decreasing the WADM insertion loss.

To summarize, we have introduced a novel design for DLSPPW-based wavelength add-drop multiplexer and showed its practical realization at telecom wavelengths. The functional device features a very compact size of $\sim20~\mu$m, low insertion loss, and reasonable wavelength selectivity allowing channel spacing of $\sim50-70$~nm in a broadband data stream.

The authors acknowledge the financial support from the Danish Agency for Science, Technology and Innovation of the project Active Nano Plasmonics (ANAP, FTP project No.~09-072949) and from the European Research Council, Grant No.~341054 (PLAQNAP).



\begin{thebibliography}{99}

\bibitem{1}T. W. Ebbesen, C. Genet, and S. I. Bozhevolnyi, Phys. Today \textbf{61,} 44 (2008).
\bibitem{2}J. Takahara, S. Yamagishi, H. Taki, A. Morimoto, and T. Kobayashi, Opt. Lett. \textbf{22,} 475--477 (1997).
\bibitem{3}M. Quinten, A. Leitner, J. R. Krenn, and F. R. Aussenegg, Opt. Lett. \textbf{23,} 1331--1333 (1998).
\bibitem{4}I. V. Novikov and A. A. Maradudin, Phys. Rev. B \textbf{66,} 035403 (2002).
\bibitem{5}T. Holmgaard and S. I. Bozhevolnyi, Phys. Rev. B \textbf{75,} 245405 (2007).
\bibitem{6}P. Berini, Phys. Rev. B \textbf{61,} 10484--10503 (2000).
\bibitem{7}V. S. Volkov, Z. Han, M. G. Nielsen, K. Leosson, H. Keshmiri, J. Gosciniak, O. Albrektsen, and S. I. Bozhevolnyi, Opt. Lett. \textbf{36,} 4278--4280 (2011).
\bibitem{13}G. Giannoulis, D. Kalavrouziotis, D. Apostolopoulos, S. Papaioannou, A. Kumar, S. I. Bozhevolnyi, L. Markey, K. Hassan, J.-C. Weeber, A. Dereux, M. Baus, M. Karl, T. Tekin, O. Tsilipakos, A. K. Pitilakis, E. E. Kriezis, K. Vyrsokinos, H. Avramopoulos, and N. Pleros, IEEE Photon. Technol. Lett. \textbf{24,} 374--376 (2012).
\bibitem{14}J. Gosciniak, L. Markey, A. Dereux, and S. I. Bozhevolnyi, Opt. Express \textbf{20,} 16300-–16309 (2012).
\bibitem{15}A. Kumar, J. Gosciniak, V. S. Volkov, S. Papaioannou, D. Kalavrouziotis, K. Vyrsokinos, J.-C. Weeber, K. Hassan, L. Markey, A. Dereux, T. Tekin, M. Waldow, D. Apostolopoulos, H. Avramopoulos, N. Pleros, and S. I. Bozhevolnyi, Laser Photon. Rev. \textbf{7,} 938--951 (2013).
\bibitem{8}T. Holmgaard, Z. Chen, S. I. Bozhevolnyi, L. Markey, A. Dereux, A. V. Krasavin, and A. V. Zayats, Appl. Phys. Lett. \textbf{94,} 051111 (2009).
\bibitem{9}Z. Chen, T. Holmgaard, S. I. Bozhevolnyi, A. V. Krasavin, A. V. Zayats, L. Markey, and A. Dereux, Opt. Lett. \textbf{34,} 310--312 (2009).
\bibitem{10}J. Gosciniak, S. I. Bozhevolnyi, T. B. Andersen, V. S. Volkov, J. Kjelstrup-Hansen, L. Markey, and A. Dereux, Opt. Express \textbf{18,} 1207--1216 (2010).
\bibitem{11}J. Grandidier, G. C. des Francs, S. Massenot, A. Bouhelier, L. Markey, J.-C. Weeber, C. Finot, and A. Dereux, Nano Lett. \textbf{9,} 2935--2939 (2009).
\bibitem{12}C. Garcia, V. Coello, Z. Han, I. P. Radko, and S. I. Bozhevolnyi, Opt. Express \textbf{20,} 7771--7776 (2012).
\bibitem{16}J. Grandidier, S. Massenot, G. Colas des Francs, A. Bouhelier, J.-C. Weeber, L. Markey, A. Dereux, J. Renger, M. U. Gonzlez, R. Quidant, Phys. Rev. B \textbf{78,} 245419 (2008).
\bibitem{17}T. Holmgaard, S. I. Bozhevolnyi, L. Markey, A. Dereux, A. V. Krasavin, P. Bolger, and A. V. Zayats, Phys. Rev. B \textbf{78,} 165431 (2008).
\bibitem{18}S. Jett\'e-Charbonneau, R. Charbonneau, N. Lahoud, G. Mattiussi, and P. Berini, Opt. Express \textbf{13,} 4674--4682 (2005).
\bibitem{19}S. I. Bozhevolnyi, A. Boltasseva, T. S\o{}ndergaard, T. Nikolajsen, and K. Leosson, Opt. Commun. \textbf{250,} 328--333 (2005).

\end{thebibliography}
 \end{document}